\documentclass[conference]{IEEEtran}
\IEEEoverridecommandlockouts
\usepackage{cite}
\usepackage{amsmath,amssymb,amsfonts}
\usepackage{algorithmic}
\usepackage{graphicx}
\usepackage{textcomp}
\usepackage{xcolor}
\usepackage{url}
\usepackage{comment}
\usepackage{listings}
\usepackage{verbatim}
\usepackage{multicol}
\usepackage{hyperref}

\def\BibTeX{{\rm B\kern-.05em{\sc i\kern-.025em b}\kern-.08em
    T\kern-.1667em\lower.7ex\hbox{E}\kern-.125emX}}
\begin{document}

\title{WISE: Unraveling Business Process Metrics with Domain Knowledge\\
}

\author{\IEEEauthorblockN{Urszula Jessen}
\IEEEauthorblockA{\textit{Dept. of Mathematics \& Computer Science} \\
\textit{Eindhoven University of Technology, the Netherlands}\\
\href{mailto:u.a.jessen@tue.nl}{u.a.jessen@tue.nl}}
\and
\IEEEauthorblockN{Dirk Fahland}
\IEEEauthorblockA{\textit{Dept. of Mathematics \& Computer Science} \\
\textit{Eindhoven University of Technology, the Netherlands}\\
\href{mailto:d.fahland@tue.nl}{d.fahland@tue.nl}}
}
\maketitle

\begin{abstract}


Anomalies in complex industrial processes are often obscured by high variability and complexity of event data, which hinders their identification and interpretation using process mining. To address this problem, we introduce WISE (Weighted Insights for Evaluating Efficiency), a novel method for analyzing business process metrics through the integration of domain knowledge, process mining, and machine learning.

The methodology involves defining business goals and establishing Process Norms with weighted constraints at the activity level, incorporating input from domain experts and process analysts. Individual process instances are scored based on these constraints, and the scores are normalized to identify features impacting process goals.

Evaluation using the BPIC 2019 dataset and real industrial contexts demonstrates that WISE enhances automation in business process analysis and effectively detects deviations from desired process flows. While LLMs support the analysis, the inclusion of domain experts ensures the accuracy and relevance of the findings.

\end{abstract}
\begin{IEEEkeywords}
process mining, anomaly detection, process conformance
\end{IEEEkeywords}

\section{Introduction}
\label{sec:intro}
Process mining primarily aims to discover, monitor, and enhance real-world processes by extracting knowledge from information system event logs~\cite{van2012process}. In industrial contexts, companies utilize process mining to achieve business goals such as improving process efficiency or reducing process costs \cite{reinkemeyer2020process}. However, the complexity of processes with numerous variants, coupled with broad contextual data and the fact that non-technical experts are often the ultimate recipients, makes these projects challenging for many companies \cite{grisold2021adoption, eggert2022applying, badakhshan2022creating}.

Generally, process analysis involves examining and evaluating the various steps and activities within a process to identify inefficiencies, bottlenecks, and areas for improvement \cite{taymouri2021business}.
Over the last decade, process mining has developed a set of tools for process analysts, ranging from process discovery to process enhancement and conformance algorithms \cite{van2022process}. However, most methods examine processes primarily from a process flow perspective \cite{van2022process}, without considering additional contextual data or domain knowledge, focusing mainly on the main behavior. Additionally, there remains a gap in academic research in applying process mining in complex scenarios \cite{taymouri2021business, thiede2018process} and improving understandability for non-experts \cite{thiede2018process}.

In contrast, event logs typically used in industry contain not only different process flow variants but also many features and attributes. Process instances executed in a business context can exhibit different performance dependent on different performance criteria. For example, some process instances may be efficient but at the same time costly due to extensive manual labor.
To tackle this problem, software vendors offer a plethora of templates, apps, and dashboards addressing various types of processes and business problems. For example, as of this publication, Celonis\,--\,one of the market leaders in process mining\,--\,offers over 145 different apps for procurement alone, each containing dozens of dashboards, KPIs, and tables. These are often connected to hundreds of tables in enterprise systems, adding an additional layer of complexity. The vast number of features and combinations requires significant time and expertise to sift through the data to find what is relevant. Finally, root cause analysis for discovered inefficiencies is mostly performed manually. Furthermore, Process Mining often emphasizes diagnostics over necessary interventions, making it crucial for the generated insights to be actionable \cite{reinkemeyer2020process}.


In this paper, we address the problem of (1) (semi-)automatically identifying the instances of the process with the highest anomalies (inefficiencies or outliers), and (2) subsequently identifying the set of context/data features that explain them, that is, their root-causes (e.g., vendors, products, variants of the process). 

The specific challenge thereby is to (1) reliably separate 'process noise' and actual anomalous process instances with minimal effort to capture domain knowledge in a specification of desired behavior, (2) to automatically identify anomalies wrt. the desired behavior, and (3) to summarize the impact of various (anomalous) features or dimensions on process outcomes in a form consumable by non-technical experts.

\begin{figure}[t]
  \centering
  \includegraphics[width=\columnwidth]{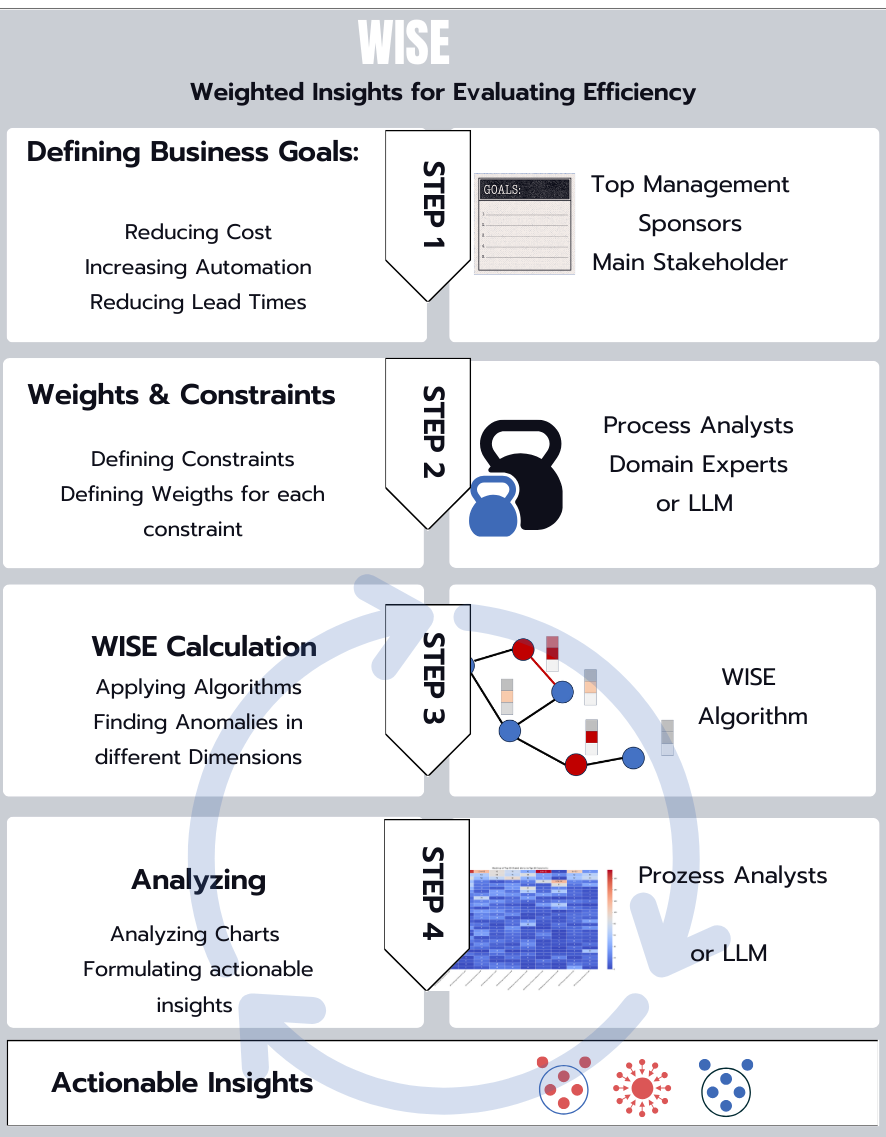}
  \caption{General WISE methodology and its main steps.} 
  \label{fig:GeneralMethodology}
\end{figure}

To address these challenges we define the WISE (Weighted Insights for Evaluating Efficiency) method, shown in 
Fig.~\ref{fig:GeneralMethodology} with the following steps:
\begin{enumerate}
    \item \textbf{Defining Business Goals}: The first step involves defining the business goals, such as reducing costs, increasing automation, and reducing lead times. The goals are usually aligned with overall company strategy and are often described by top management, project sponsors or main stakeholders. This sets the context and objectives for the subsequent analysis.
    \item \textbf{Defining Constraints and Weights}: Next, domain experts, process analysts and/or machine learning models (LLM) collaborate to define constraints and assign weights to each constraint. These constraints are critical in shaping the analysis and ensuring that the most relevant factors are considered.
    \item \textbf{Applying WISE for Analysis}: Using the WISE framework, we apply algorithms to the event logs to find anomalies and deviations in different dimensions. This involves comparing all traces to the intended flow and computing deviation statistics against each feature-value combination. Each process instance is evaluated and scored based on constraints and weights for each of previously defined objectives. 
    \item \textbf{Visualization and Insight Generation}: The scores are visualized at a fine-grained level to reveal the exact features contributing to specific performance deviations. The individual process instances scores are aggregated for different dimensions. 
    \item \textbf{Actionable Insights for Improvement}: Finally, the insights derived from the analysis are used to address the identified issues. For example, this could involve recommendations for renegotiating contracts with specific suppliers, providing targeted training for departments, or reengineering parts of the process in some specific context.
\end{enumerate}
This methodology aims to simplify the assessment of process performance indicators by making them more interpretable and actionable for professionals through simplification and approximation. By utilizing specialized industry knowledge and focusing on essential tasks, we intend to clarify these indicators for different stakeholders in process mining projects.

Next, we review the literature and describe the problem in more detail in Sect.~\ref{sec:related_work} and provide technical details of our method in Sect.~\ref{sec_method}). In Sect.~\ref{sec:evaluation}, we demonstrate our method on the BPIC 2019 dataset and report on evaluating the method in an industrial case study. Feedback from process owners  indicated that the method effectively identified issues leading to actionable insights.

\section{Related Work and Problem Description}
\label{sec:related_work}

Historically, businesses have continuously sought methods to refine their internal processes to achieve these goals. One of such methods - Business Process Analysis (BPA) emerged as a foundational approach to systematically examine and improve business processes\cite{dumas2018introduction}. Traditionally, BPA methods have been time-consuming, involving extensive meetings, document analysis, and direct observation. However those traditional methods,were not only cost-intensive, but often lead to discrepancies between actual and perceived processes, particularly in complex scenarios  \cite{stefanini2020process}. Process mining, which leverages data from software systems to discover, check conformance, and enhance processes,compared with BPA can be executed semi-automatically \cite{van2012process}. 
Process Mining Research has put a lot of effort to support industry in analyzing real life processes by developing different algorithms and methods. On one hand there is a plethora of algorithms such as Fuzzy Mining and clustering techniques simplify complex models, making them more understandable and manageable\cite{stefanini2020process}. Additionally the research on Declarative Process Mining Utilizing constraint-based languages to define flexible sequences of activities, allows for more adaptive and resilient process designs\cite{di2022declarative}.

Despite those aforementioned efforts, significant challenges, in real-world scenarios, still remain unresolved. The Process Mining Manifesto \cite{van2012process} outlines 11 of such challenges, including dealing with complex event logs and improving understandability for non-experts. Furthermore many publications identify challenges such as managing complex business processes, \cite{eggert2022applying}, securing stakeholder buy-in, and aligning expectations\cite{dakic2019process} as critical for effective process mining adoption.Similarly R’Bigui and Cho \cite{r2017state} and Munoz-Gama et al. \cite{munoz2022process}, note that, especially in the area of complexity and understandability, many of these challenges remain unresolved. Likewise Thiede et al. emphasize the need for more research on applying process mining in complex scenarios \cite{thiede2018process}.

\begin{figure}[ht]
  \centering
  \includegraphics[width=\columnwidth]{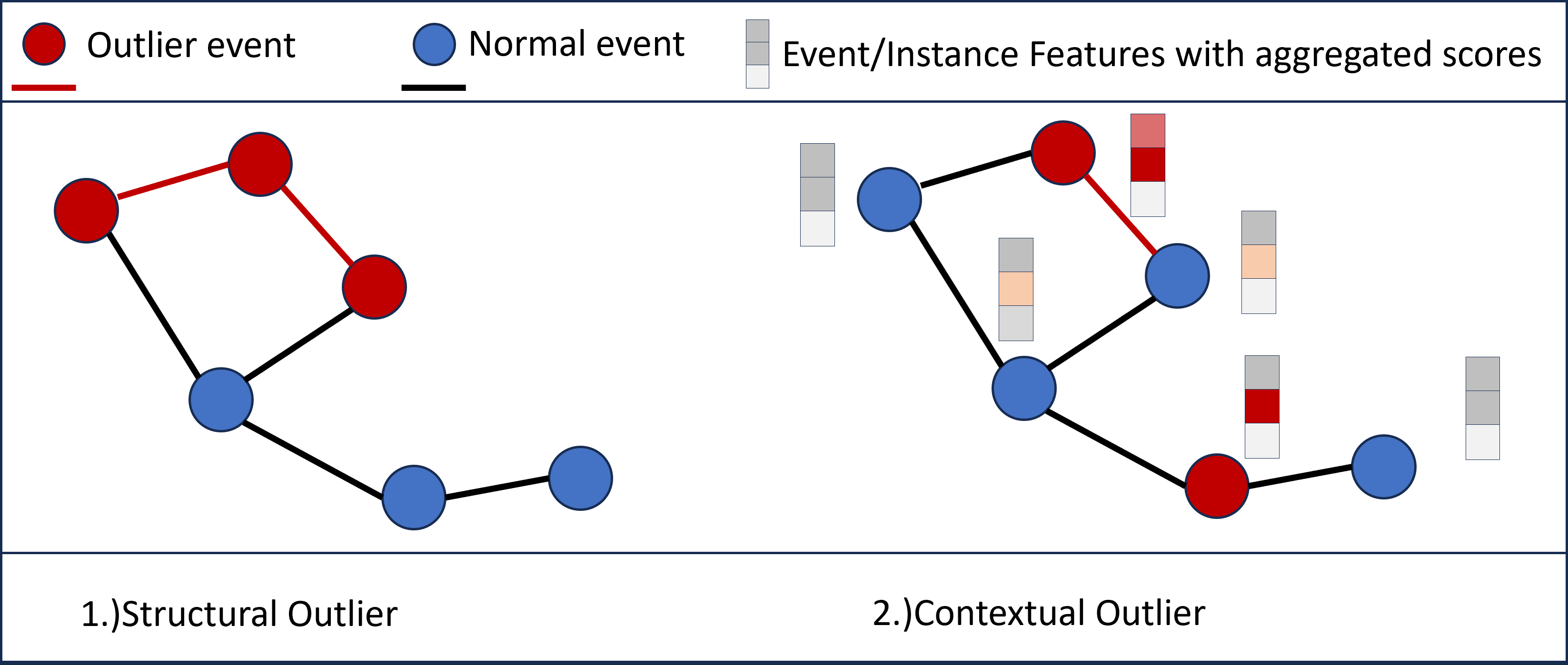}
  \caption{Structural and contextual outliers. Based on \cite{liu2022benchmarking}, modified.} 
  \label{fig:ContextualOutlier}
\end{figure}

One of the features of such complex, multi-variant processes is the presence of noise, outliers, and infrequent behavior\cite{koschmider2021demystifying}. In event logs, outliers are defined as unusual or exceptional instances that deviate from normal patterns, while noise refers to random or erroneous data points that do not follow expected behaviors \cite{koschmider2021demystifying}. A common technique for simplifying data interpretation is filtering out such instances \cite{van2020detection, sun2019filtering}. However, Koschmider et al. argue that detecting and addressing contextual outliers, which are unusual within specific contexts, and collective outliers, which are subsequences deviating from the norm even if individual activities are typical, is crucial for improving process mining accuracy and gaining valuable process insights \cite{koschmider2021demystifying}. As proposed by Liu et al.\cite{liu2022benchmarking} WISE methodology uses contextual outliers, calculating scores for event or process instance features as depicted in Fig.~\ref{fig:ContextualOutlier}.

Despite the increase in commercial software tools capable of generating dashboards and visualizations directly from core data systems like SAP, the anticipated business value may not always be achieved, potentially leading to disillusionment \cite{emamjome2019case}. The challenge in process mining involves enhancing output clarity and usability for non-expert stakeholders, especially within complex industrial settings \cite{van2012process}. Transforming complex data into understandable insights remains a well-known issue in data science, highlighting the importance of 'analytics translators'—professionals skilled at interpreting data to identify business challenges and convert needs into data-driven inquiries \cite{herring2019train, zhang2023deliberating}.

Traditional algorithms often produce overly complex, "spaghetti-like" models in unstructured processes, complicating interpretation \cite{stefanini2020process}. Successful process mining projects require methodologies that integrate domain knowledge, set clear process goals, encourage iterative analysis, and provide clear, practical guidelines for each project stage \cite{van2015pm}.

Process mining analysts play a crucial role by obtaining and preparing data, analyzing processes, generating insights, and providing recommendations for improvement \cite{zimmermann2022process}. Zimmerman et al. describes the biggest challenges perceived by process analysts, such as understanding complex process domains and terminology, collaborating with stakeholders, handling the inherent complexity of real-world processes, and ensuring proper training and tool knowledge. These issues underscore the need for better support and solutions to enhance the effectiveness of process mining initiatives \cite{zimmermann2022process, zerbato2021initial, zimmermann2023makes}.

Zimmermann et al. argue that involving stakeholders is a critical strategy to address these challenges. This approach aims to increase understanding, align expectations, and identify key requirements, involving executives, business owners, process owners, IT, and domain experts. Such involvement ensures accurate data interpretation, effective communication, and the development of relevant solutions \cite{zimmermann2023makes}.

The WISE methodology aims to address described challenges of analyzing processes in complex environments at the same time delivering the interpretable process insights. Using contextual and collective outliers\cite{koschmider2021demystifying} the method on approximation
and summarization to distinguish between mere ’process noise’ and
genuine insights that provide a foundation for actionable decisions.

\section{Proposed method: WISE}
\label{sec_method}

Our proposed WISE method is grounded on the business objectives defined by stakeholders, managers and domain experts (cf. Sect~\ref{sec:intro}). 

At the heart of our methodology is the development of a \emph{Process Norm} (Sect.~\ref{sec_method:norm}). Technically, the norm is a set of (very simple) constraints stating desired behaviors. It serves as a reference point, facilitating the distinction between routine data variations\,--\,commonly seen as noise\,--\,and authentic anomalous activities that merit further analysis for enhanced process understanding. 

We then automatically measure the degree of deviation of each process execution from the constraints in the Norm through weighted scoring functions (Sect.~\ref{sec_method:scoring}. Deviations scores of all process executions can then be aggregated wrt. individual features (e.g., vendors, products) to measure which feature (e.g., Vendor A vs Vendor B) is associated with how much deviation. We then show in Sect.~\ref{sec:evaluation} that this allows to derive insights by process analyst or by LLMs. If necessary the WISE analysis can be repeated only for selected features and data.


\subsection{Definition of Process Norm}
\label{sec_method:norm}

Initially, we establish the Generalized Process Norm as a comprehensive, multi-layered framework that incorporates essential events and attributes reflective of typical process operations. This norm defines a series of constraints for subsetting the event log, with constraints and weights specified for each business objective, referred to as a view. Each view can be analyzed independently or as part of multiple perspectives on the process. The constraints are organized into different layers, each focusing on a specific aspect:

\begin{itemize}
    \item \textbf{Layer 1}: Mandatory events (without imposing order)
    \item \textbf{Layer 2}: Mandatory ordering of events
    \item \textbf{Layer 3}: Balanced events - events that should occur the same number of times
    \item \textbf{Layer 4}: Singular events - events that should not occur more than once
    \item \textbf{Layer 5}: Exclusive events - events that should not occur at all
\end{itemize}
More layers can be added depending on analysis goals.
This structure allows for a detailed and flexible analysis of business processes according to various objectives.

Process Norms grounds all process instances and enable discovering deviations and inefficiencies in process flow.
We assume as input an event log \(L = (E, \#, \prec)\), comprising events \(E \subseteq U_{\text{ev}}\) which carry attributes by \(\# \in E \rightarrow U_{\text{map}}\) as attribute-value pairs $U_{\text{map}} : U_{\text{attr}} \to U_{\text{val}}$), and a strict partial ordering of events \(\prec \subseteq E \times E\) establishes a strict partial ordering of events.
We assume attributes \texttt{activity}, \texttt{case} $\in U_{\text{attr}}$ with their standard interpretation. For a case identifier $c$, the trace is the partial order $\sigma_c \in L = (E_c,\#_c,< _c)$ with events $E_c = \{ e \in E \mid \#_{\text{case}}(e) = c \}$ and $\#_c = \#|_{E_c}$ and $< _c = <|_{E_c \times E_c}$ restricted to $E_c$\cite{van2022process}. Note that this definition includes ``classical'' sequential event logs.

To identify anomalous process executions, we define a ``Process Norm'' stating desired behavior. With the aim to allow domain experts with no formal training in process modeling to specify the norm, we turn to simple constraints over events in the log. Drawing on declarative process modeling languages such as DCR graphs\cite{di2022declarative}, we limited the process norm only to most basic constraints as these are easy to understand and specify.

\textbf{Generalized Process Norm}:
\label{def:process_norm}
A \textit{Generalized Process Norm} \(N =  (CR_N, W_N)\) defines a series of constraints \(CR_N = \langle c_1, c_2, \ldots, c_n\rangle = \langle M, <_M, H, U, A\rangle \) (grounded in business understanding) specifying:
    \begin{itemize}
        \item Mandatory Activities: \(M \subseteq U_{\text{activity}}\) is the set of mandatory activities.
        \item Sequential Directives: \(<_M \subseteq M \times M\) is the set of pairs of activities where the second activity must eventually follow the first.
        \item Equilibrium Standards: \(H \subseteq 2^{U_{\text{activity}}}\) are sets of activities that should occur in balanced amounts.
        \item Singularity Criteria: \(U \subseteq U_{\text{activity}}\) is the set of unique activities that should not occur more than once per process instance.
        \item Exclusion Guidelines: \(X \subseteq U_{\text{activity}}\) is the set of undesirable activities that should not occur at all.
    \end{itemize}
We call each $c_i$ also a \emph{layer} of the norm. The weights \(W_N = \langle w_1, w_2, \ldots, w_n \rangle, w_i \in [0, 1]\) specify the importance of the constraint \(\langle c_1, c_2, \ldots, c_n\rangle\) in the overall calculation.


In WISE, domain experts define the norm by selecting the process constraints in each layer and defining their weight. In absence of domain experts, constraints and corresponding weights can be also generated automatically through a contextualized Large Language Model, e.g., by prompting for mandatory activities in a specific type of procurement process.

\subsection{Scores Calculation}
\label{sec_method:scoring}

Building on the multi-layered framework for defining process norms, we differentiate marginal differences from actual anomalies by scoring each individual process instance $\sigma_i$ with events $E_\{\sigma_i\}$ based on their deviations from the norm. 
Penalties for each violation are calculated based on the constraints $c_j$ and the weight $w_j$ in each layer.

\textbf{General Penalty Calculation} For each constraint \(c_j \in CR_N\) with weight \(w_j\), define a penalty based on the degree of violation of the constraint. The \emph{penalty for a process instance} \(i\) is calculated as:
    \begin{equation} \label{eq_penalty}
    penalty_i = \sum_{j=1}^{n} w_j \cdot f(c_j, \sigma_i)
    \end{equation}
where \(f(c_j, \sigma_i)\) is a function that measures the degree of violation of the constraint \(c_j\) in the trace \(\sigma_i\). The adjusted score for the process instance \(S_i\) is then:
    \begin{equation} \label{eq_score}
    S_i = 1 - penalty_i
    \end{equation}
To ensure that the score is not smaller than 0 and to normalize it, we define the \emph{normalized score} as:
\begin{equation} \label{eq_normalized_score}
S_i^{\text{norm}} = \frac{S_i}{S_{\text{max}}} \quad \text{if } S_{\text{max}},{S_i} > 0 \text{ else } 0
\end{equation}
where \(S_{\text{max}}\) is the total possible score.

We now define the penalty functions $f(c_i,.)$ for each layer of the process norm.

\textbf{Layer 1: Foundational Protocols:}
    To evaluate adherence to the process norm, we identify the constraints \(CR = \{(A, w_A), (D, w_D), (G, w_G)\}\) where \(A, D, G \in \Sigma\) are activities and \(w_A, w_D, w_G\) are their respective weights.
    For a trace \(\sigma_i\) with events \(E_{\sigma_i} = [A, D, A, F, C, \ldots]\), we check if all constraints are satisfied. If all activities \(A, D, G\) are present, the score for this layer \(S_i\) is 1. If any constraint is not satisfied, we add a penalty based on the weight of the missing activity.
    Score Calculation for \(M\) (Mandatory Activities):
    \begin{equation} \label{eq_mandatory}
    f(M, \sigma) = \sum_{a \in M} (1 - \delta(a \in E_{\sigma}))
    \end{equation}
    where \(\delta\) is an indicator function that is 1 if the activity \(a\) is present in \(E_{\sigma}\) and 0 otherwise.

\textbf{Layer 2: Sequential Directives:}
    For each pair \((a_1, a_2) \in <_M\), we check if \(a_1\) eventually follows \(a_2\) in the trace \(\sigma_i\). We define $seq(a_1,a_2,\sigma) = \delta(\exists (e_1, e_2) \in E_{\sigma} : \#_{\text{activity}}(e_1) = a_1, \#_{\text{activity}}(e_2) = a_2, e_1 <_{\sigma} e_2))$ which returns 1 if the condition holds and 0 otherwise. Then the penalty for \(<_M\) is:
    \begin{equation} \label{eq_sequential}
    f(<_M, \sigma) = \sum_{(a_1, a_2) \in <_M} (1 - seq(a_1,a_2,\sigma))
    \end{equation}
    Note that (\ref{eq_sequential}) gives a penalty of 1 for each pair $(a_1,a_2) \in <_M$ that is \emph{not} present in $\sigma$.

\textbf{Layer 3: Equilibrium Standards:}
    For each equilibrium set \( \{h_1,\ldots,h_k\} \in H\), we check if all activities occur equally often in $\sigma_i$. Let $\text{count}(h_r, \sigma) = |\{ e \in E_\sigma \mid \#_{activity}(e) = h_r\}|$. Taking one activity $h_1$ as reference point, we penalize for every $h_r,r>1$ occurring more/less often than $h_1$:
    \begin{equation} \label{eq_equilibrium}
    f(H, \sigma) = \sum_{\{h_1,\ldots,h_k\} \in H}\sum_{r=2}^k |\text{count}(h_r, \sigma) - (h_1, \sigma)|
    \end{equation}

\textbf{Layer 4: Singularity Criteria:}
    For each event \(u \in U\), we penalize if the event occurs more than once in the trace \(\sigma_i\):
    \begin{equation} \label{eq_singularity}
    f(U, \sigma) = \sum_{u \in U} \max(0, \text{count}(u, \sigma) - 1)
    \end{equation}

\textbf{Layer 5: Exclusion Guidelines:}
    For each event \(x \in X\), we penalize if the event occurs in the trace \(\sigma_i\):
    \begin{equation} \label{eq_exclusion}
    f(X, \sigma) = \sum_{x \in X} \text{count}(y, \sigma)
    \end{equation}


This approach scores each process instance against a structured set of norms, weighted by importance.

\section{Evaluation and Results}
\label{sec:evaluation}

We implemented WISE in a Python library. We demonstrate its application on the BPI 2019 Challenge (data and analysis goal) and report on a case study that applied WISE to industrial dataset with the aim of receiving feedback for its effectiveness in identifying process inefficiencies and corresponding features. Code and results are available at~\url{https://github.com/feelfine1977/WISE}.

\subsection{Evaluation of the BPI 2019 Dataset}
We analyzed the BPI Challenge 2019 dataset, sourced from a multinational corporation in the coatings and paints sector in The Netherlands. It records in detail the purchase order handling process across its 60 subsidiaries, with a specific focus on compliance \cite{vanDongen2019BPIDataset}.

\begin{figure}[t]
  \centering
  \includegraphics[width=\columnwidth]{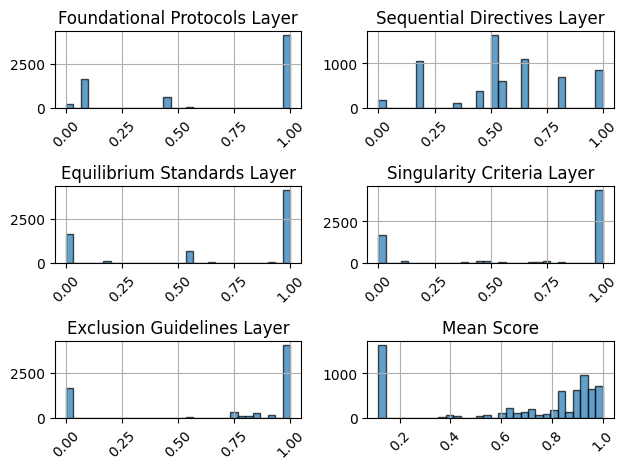}
  \caption{Scores distribution in BPIC 2019 Dataset based on the defined Process Norm.} 
  \label{fig:ScoresDistributionBPIC2019}
\end{figure}

The challenge states 4 categories of purchase order line items (indicated by absence or presence of flags in related activities); each category should follow a distinct process flow, see~\cite{vanDongen2019BPIDataset} for details.

Using the method outlined in Sect.~\ref{sec_method}, we defined a process norm for ``Process Standardization'' based the 1st author's comprehensive understanding of procurement operations and insights from the BPIC 2019 description and reports. For instance, essential activities for every procurement cycle, such as "Create Purchase Order Item" and "Record Goods Receipt," were included in the \textit{Foundational Protocols Layer}. Contrarily, activities like "Change Price" and "Change Vendor" were identified as counterproductive due to their manual nature, introducing inefficiencies, and were thus added to the \textit{Exclusion Guidelines Layer} constraints.

We scored all process executions in the log according to this norm; Fig.~\ref{fig:ScoresDistributionBPIC2019} shows the distribution of scores for different layers. The foundational protocols layer reveals the presence of 4 process flow. This distribution in the other layers suggests that different types of process flows may require different types of constraints. 

To show that, depending on the business objective, scores can vary significantly, we adjusted the weights in the norm to fit a ``Process Cost'' view. Tab.~\ref{table:process_data} shows the different scores for each layer in the different views for a single process instance. Here, we observe high process scores for the \textit{Foundational Layer} from the \textit{Standardization Process View} perspective, while the score for the \textit{Process Cost View} perspective is much lower. 

\begin{figure}[t]
  \centering
  \includegraphics[width=\columnwidth]
  {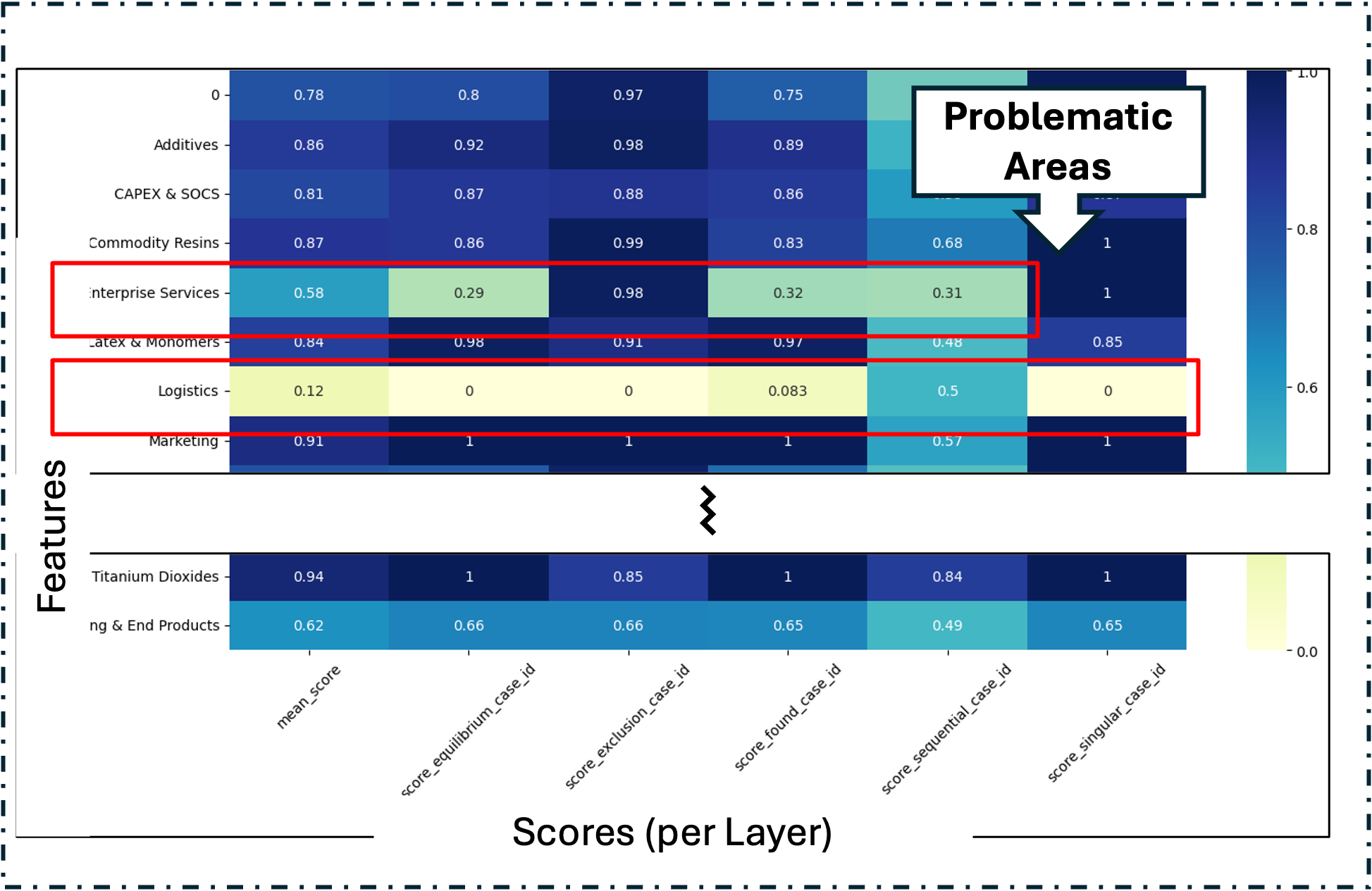}
  \caption{Scores Distribution for Data from previous dataset with filter only on those with low performing scores.} 
  \label{fig:Score_Logistic}
\end{figure}

To identify ares of high inefficiencies in the entire process, we aggregate the scores of all process executions. Fig.~\ref{fig:Score_Logistic} shows the scores of all layers (x-axis) wrt. the features in the event log (y-axis) in a heatmap. We can compare the score distribution all event log features. The coloring shows at first glance that cases in the \textit{Logistic} category are particularly impacted by low scores, directing the analyst in their search for explanations.


\begin{table}[t]
\centering
\resizebox{\columnwidth}{!}{
\begin{tabular}{|l|c|c|}
\hline
Layer & View Process Standardisation & View Process Costs \\ \hline
Foundational Layer & 0.95 & 0.50 \\ \hline
Sequential Layer & 0.85 & 0.85 \\ \hline
Equilibrium Layer & 0.75 & 0.75 \\ \hline
Singularity Layer & 0.55 & 0.75 \\ \hline
Exclusion Layer & 0.60 & 0.85 \\ \hline
\end{tabular}}
\caption{Scores for one process instance over different layers with two views: process standardisation and process costs.}
\label{table:process_data}
\end{table}

Based on findings depicted in Figure \ref{fig:Score_Logistic} we generated another heatmap for scores over individual values in for \textit{Logistic}. There, we identified particular types of low performers such as \textit{Road Packed}, \textit{Sea} and \textit{Warehousing} which performed exceptionally poorly. We could also discover that areas of \textit{Exclusion Layer}, \textit{Equilibrium Layer} and \textit{Singularity Layer} showed poorest score, which could suggest that in those areas a lot of manual changes and rework was taking place. This could be identified as possible root cause for the problems in this particular areas. Apart from that we could also discover poor performance in \textit{Service} Category and interesting patterns of particular Vendors.

\subsection{Comparing BPIC 2019 Reports to WISE Reporting}

\begin{figure}[t]
  \centering
  \includegraphics[width=\columnwidth]{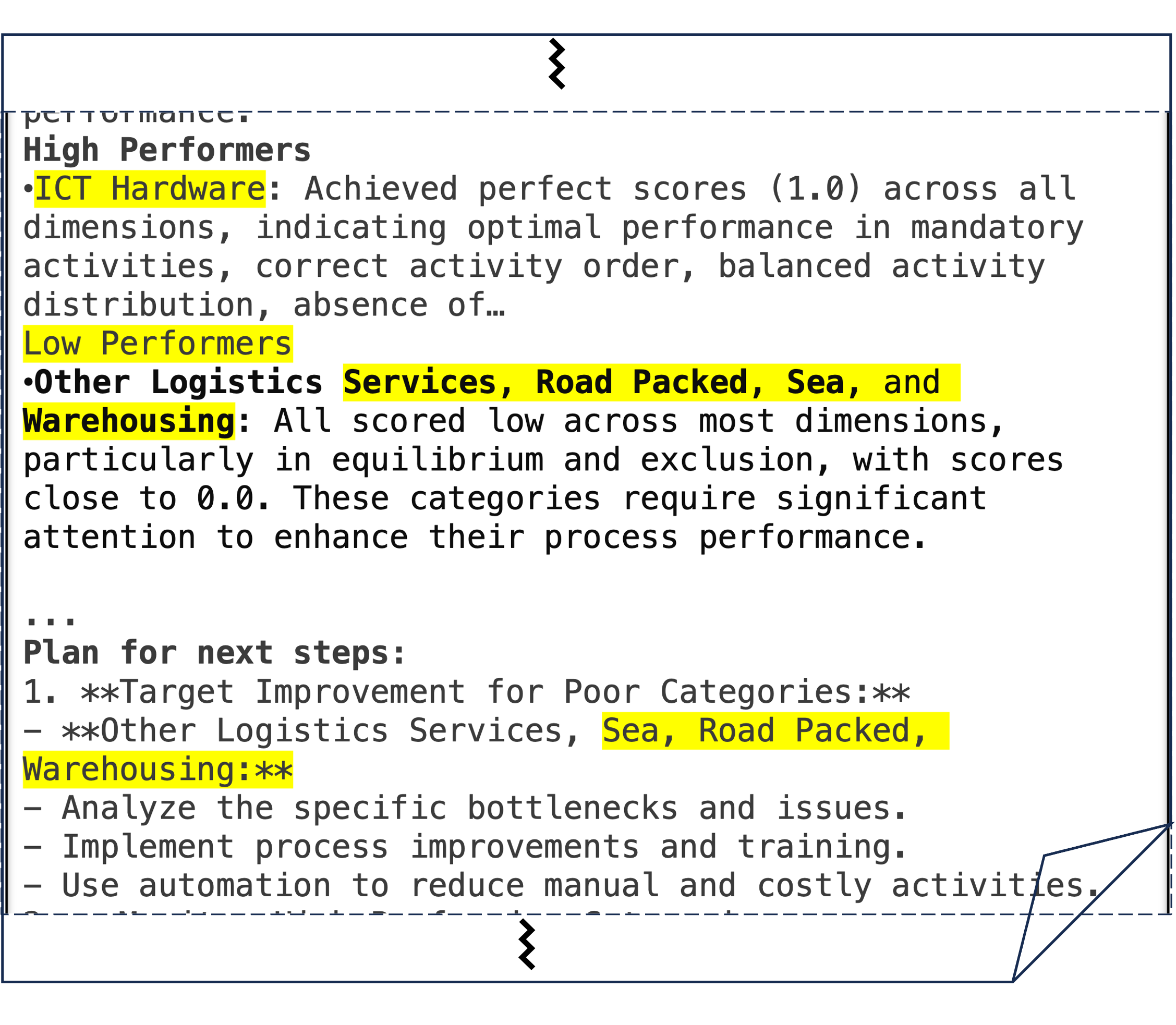}
  \caption{Part of LLM \textit{ChatGPT4} analysis suggesting additional areas of interest and further steps.} 
  \label{fig:ChatGPT}
\end{figure}

The WISE method not only facilitates various data dicing, slicing, and drilling-down techniques combined with statistical and machine learning methodologies but also enables the automated detection of features that have significant impact on process performance. Our automated workflow allows for utilizing Large Language Models in the role of the analyst. This can be used to support Process Norm creation, add weights to different layers, analyze singular figures, or even create whole reports with management summaries and detailed project plans for process improvement in the form of PDF files.

\begin{figure}[t]
  \centering
  \includegraphics[width=\columnwidth]{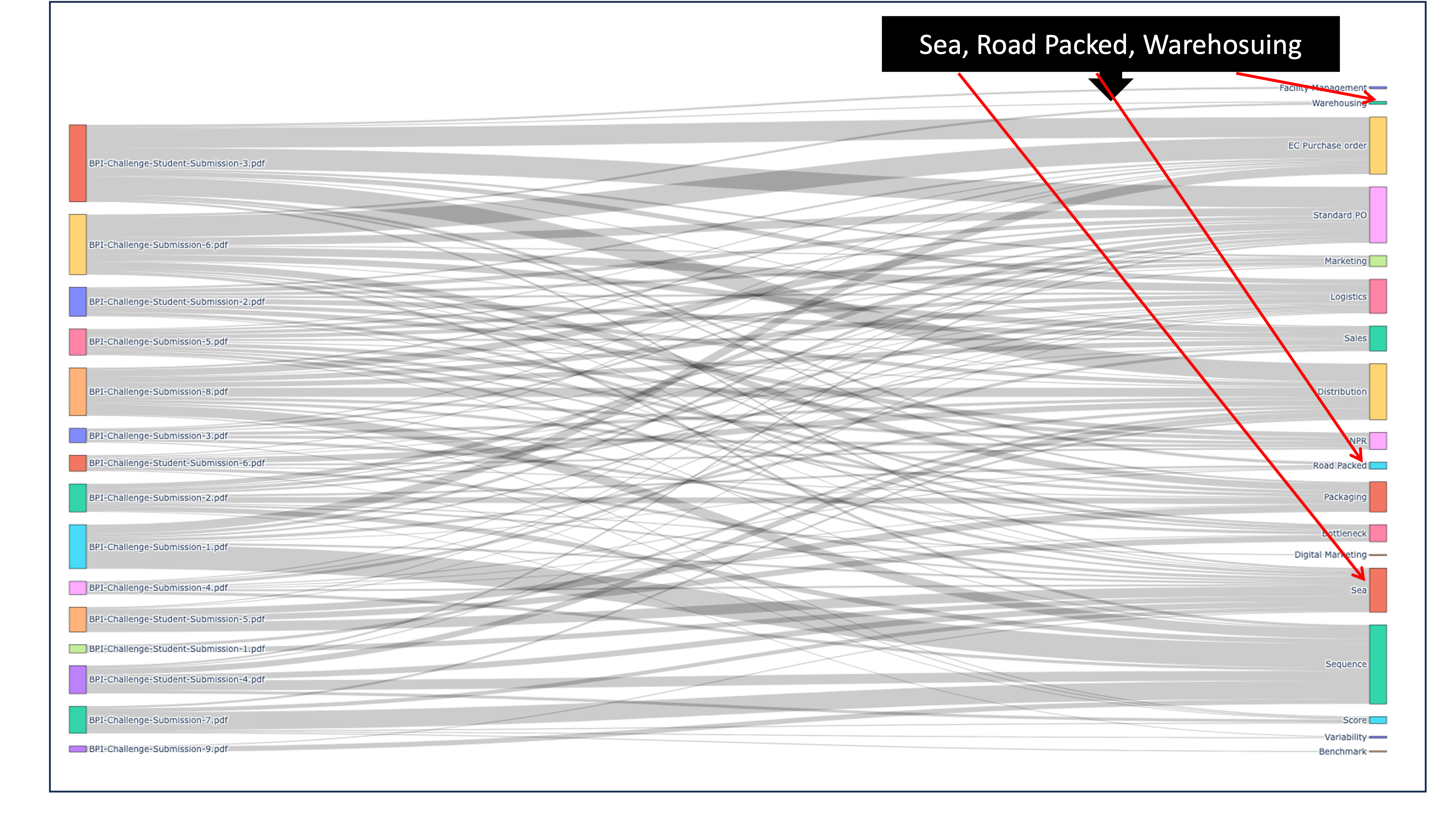}
  \caption{Shared words between BPIC 2019 Challenge Papers and automated Report.} 
  \label{fig:SharedWords}
\end{figure}
For a more comprehensive analysis, we created a detailed report (74 pages) for the BPIC 2019 dataset, using Chat-GPT 4o as analyst, and compared it with all 15 papers presented as part of the BPIC 2019 challenge. 

Figure \ref{fig:ChatGPT} illustrates an excerpt from this automated analysis, highlighting categories with high and low performers, and suggesting further detailed investigations. We assessed how the automated report differed from the manually analyzed results in the BPIC 2019 challenge papers. 

Figure \ref{fig:SharedWords} shows the words shared between the challenge papers and the automated report. Some areas, such as packaging or marketing, are clearly identified as important in both the challenge papers and the WISE Report, as indicated in the figure. Additionally, most analyses described logistics as one of the areas requiring further examination. Only few papers explained those inefficiencies through further analysis and finding features such as Road Packed, Sea or Warehousing as depicted in Figure \ref{fig:Score_Logistic}. Additionally, the challenge papers generally did not delve deeper or analyze in detail different correlations such as specific employees patterns or vendors. The WISE analysis provided these findings almost instantly.

Overall, the challenge papers were more focused on process flow attributes, whereas the WISE Report identified the most substantial areas for improvement, offering detailed proposals. These detailed findings, compounded with different layers representing various types of challenges, enabled the formulation of hypotheses about root causes, such as the incorrect order of mandatory activities or rework. However the automated analysis was sometimes too detailed and summary was often not concise enough. Additionally the proposed improvements seemed to be often too generic. This proves that although such automated analysis can be helpful and time efficient, the human in the loop is still necessary to deliver reports that are exactly aligned with stakeholders expectations.

\subsection{Industrial Case Evaluation}

The method was also evaluated in an industrial setting within a real estate company managing a substantial property portfolio. This evaluation was part of a procurement improvement initiative aimed at enhancing the transparency and standardization of the procurement process, which is crucial for effective property management. While the classical process analysis revealed several bottlenecks and inefficiencies, it struggled to pinpoint specific issues due to the overwhelming data, very high variability and unclear root causes.

The optimization project faced several challenges:
\begin{enumerate}
    \item \textbf{Supply Chain Variability}: The vast number of suppliers introduces variability in quality, reliability, and terms of supply, leading to inconsistencies in procurement practices and challenges in maintaining standardization.
    \item \textbf{Materials Management Complexity}: Managing over 100,000 different materials adds substantial complexity, as different materials may require varied handling and procurement strategies.
    \item \textbf{Quality Assurance Issues}: Ensuring that all ordered materials meet specified quality standards is critical. Robust quality assurance processes are essential to prevent the use of substandard materials, which can cause maintenance failures and increased costs.
\end{enumerate}

Addressing these challenges, the project aimed to enhance transparency and standardization in the procurement process. One main objective was to identify drivers of variability, determining whether variability is driven by material properties, specific suppliers, or inherent process complexities.

\subsubsection{Implementation of Scoring Mechanism}

\begin{figure}[t]
  \centering
  \includegraphics[width=\columnwidth]{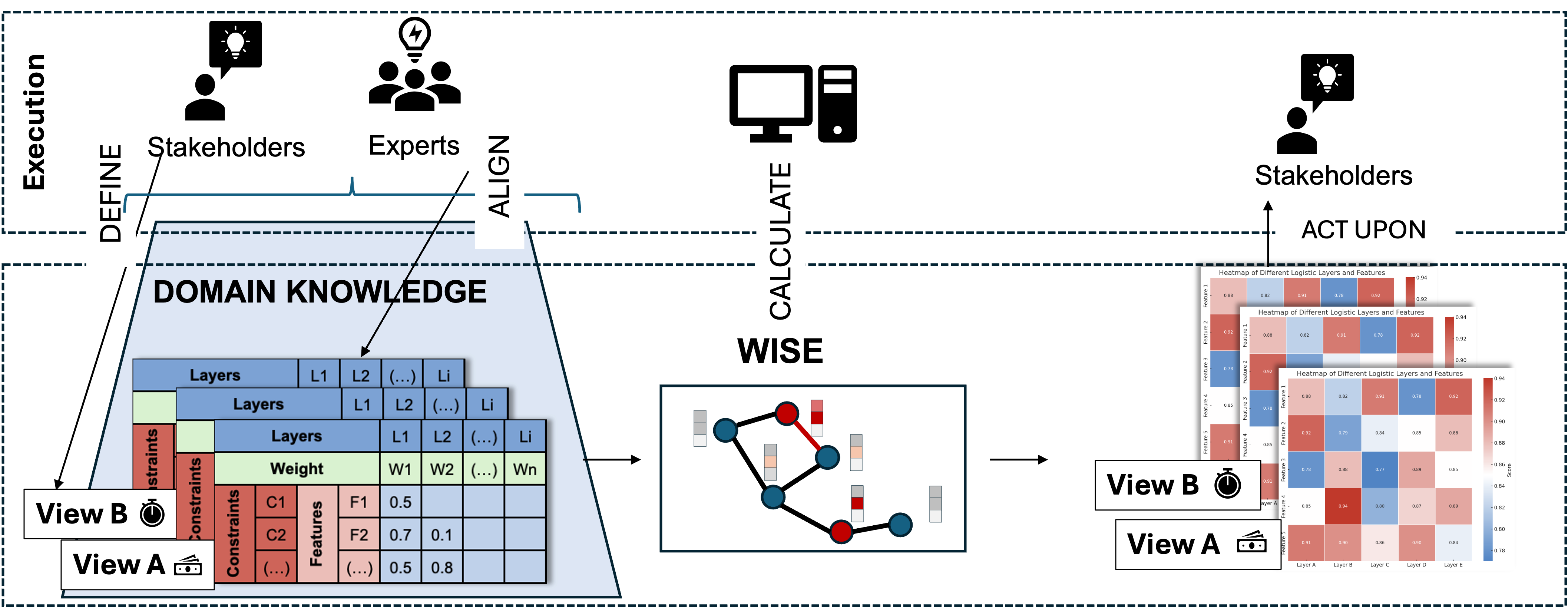}
  \caption{Phases of the process improvement project supported by the WISE methodology.}
  \label{fig:WISEMethod}
\end{figure}

The business objective of decreasing supply chain variability was defined as part of a broader strategy through top management. Grounded on this goal, we conducted internal procurement workshops to develop a matrix that evaluates the norms and the significance of individual problems identified within the process. Figure \ref{fig:WISEMethod} depicts different phases of the project and the parties responsible for various stages of execution.

Domain experts and process analysts were invited to the project to ensure the Process Norm aligned with the defined business objectives. Discussions about constraints and their weights focused on actionable areas within the company's control. The development of the Process Norm was significantly faster than typical process analysis workshops, as it was unnecessary to design all possible constraints. Instead, the focus was on creating constraints that could suggest actionable interventions, such as renegotiating specific contracts or designing new ERP templates for particular types of materials.

Based on the Definition of the Process Norm defined in \ref{def:process_norm}, we defined the constraints and weights for all layers. The following are excerpts of findings for particular layers for the specific business objective:

\begin{itemize}
    \item \textbf{Sequential Directives}: Simplified but included variants addressing processes under a Framework Agreement, eliminating the need for steps like additional approvals.
    \item \textbf{Equilibrium Standards}: Critical for process instance variability, verifying quantities, quality checks upon goods receipt, and financial consistency from purchase requisition to invoice payment.
    \item \textbf{Singularity Criteria}: Aimed to eliminate repeated releases and checks, identified as significant contributors to increased process time and variability.
    \item \textbf{Exclusion Guidelines}: Identified costly and detrimental tasks to process automation, including unnecessary communications and changes to critical elements such as prices, quantities, currency, or key fields in purchase orders.
\end{itemize}

\subsubsection{Results of Procurement Optimization Project}
\begin{figure}[t]
  \centering
  \includegraphics[width=\columnwidth]{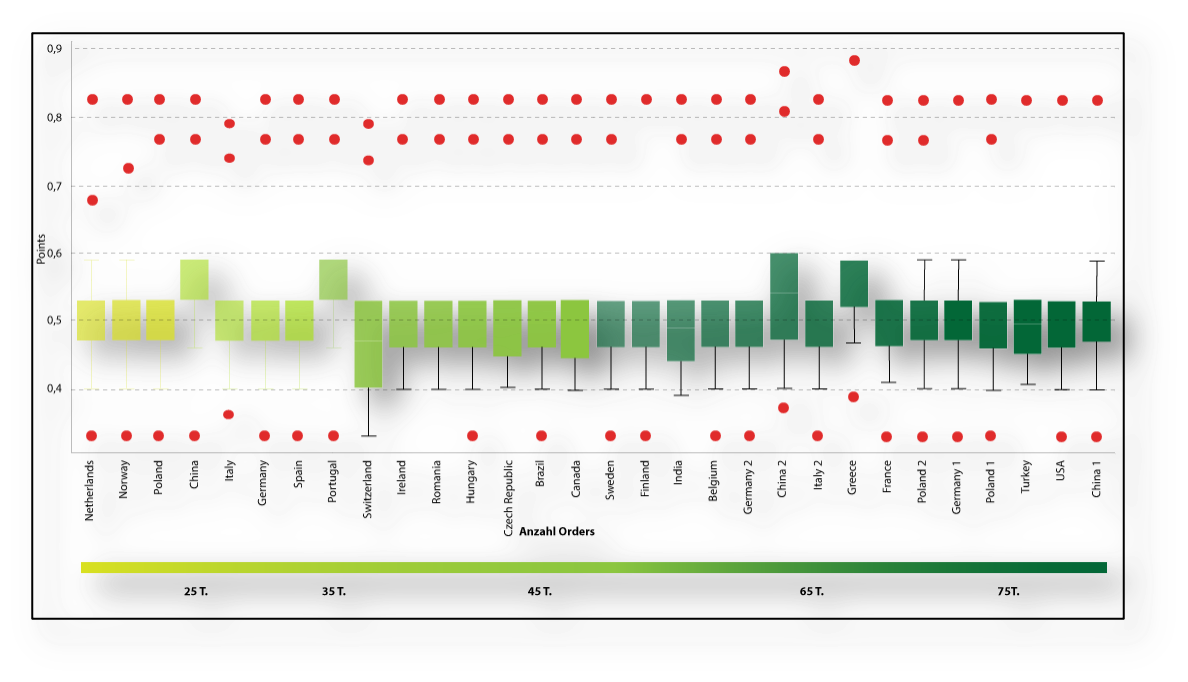}
  \caption{Example of Summarized Norms Scoring grouped by different locations.}
  \label{fig:AllNorms1}
\end{figure}

\begin{figure}[t]
  \centering
  \includegraphics[width=\columnwidth]{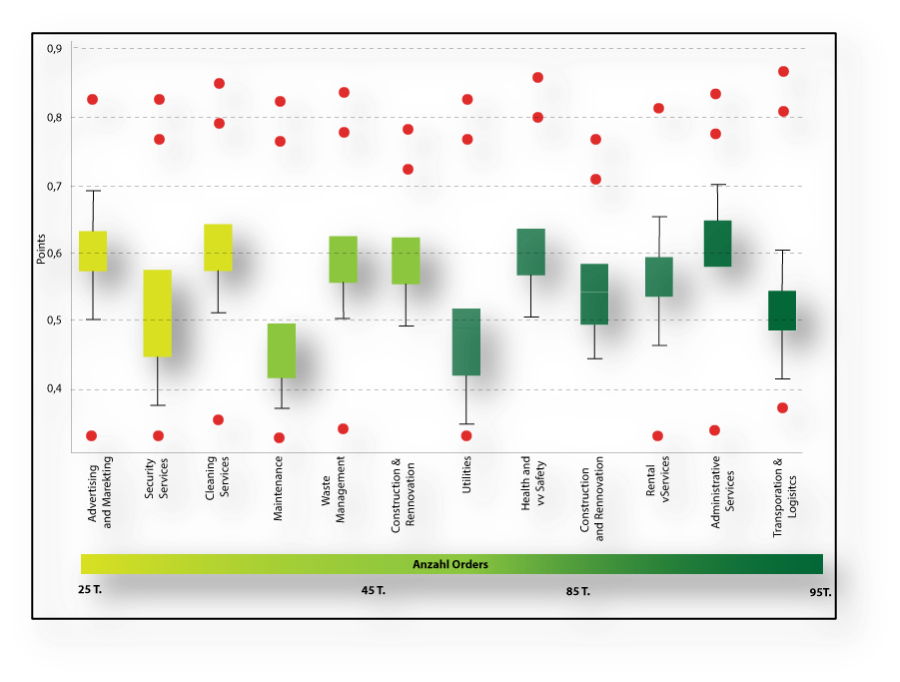}
  \caption{Example of Summarized Norms Scoring based on material group.}
  \label{fig:AllNorms2}
\end{figure}

The analysis phase of procurement optimization project has revealed various areas of improvement and facilitated additional communication between different project stakeholders.
Figures \ref{fig:AllNorms1} and \ref{fig:AllNorms2} depict as an example some of the anonymized results of the analysis.
Figure \ref{fig:AllNorms1} synthesizes a detailed assessment of scoring norms across properties, grouped by geographic location. The boxplot segments, whose darker color means the biggest volume of orders, correspond to scores calculated by WISE algorithms.
One of the findings was that the though small variability in compliance is visually evident, those differences does not support location as variability indicator. Additionally a higher volume of orders does not directly correlate with lower process scores.
Figure \ref{fig:AllNorms2} provides on the other hand an analysis based on the aggregated scores for each material group. In contrast to Figure \ref{fig:AllNorms1}, which showed a more homogeneous distribution of scores, Figure \ref{fig:AllNorms2} reveals pronounced variability among different material groups, with security services, maintenance, and utilities emerging as areas of concern due to their problematic scores.

The analysis efficiently pinpointed significant challenges within various categories. Subsequent investigations revealed specific correlations, such as suppliers deviating from standard procedures or equipment requiring additional user training for departmental staff. Overall, the project successfully highlighted sectors where improvements are feasible. Recommendations include renegotiating contracts with certain suppliers, modifying material templates for select categories, and implementing specialized documentation or training for specific equipment. These strategic measures are anticipated to yield substantive improvements in the procurement process.

\section{Conclusion}

This study demonstrates the effectiveness of the WISE methodology in identifying and addressing process inefficiencies within complex procurement environments. By defining process norms and applying a structured scoring mechanism, the methodology highlights areas of variability, inefficiency, and potential improvement.

The evaluation of the BPI 2019 dataset revealed that different types of process flows require distinct constraints and norms, enabling more precise tailoring of process improvement initiatives. The detailed scoring of process instances provides a comprehensive view of process performance and areas for enhancement.
The industrial case study emphasizes the practical applicability of the WISE methodology.One of the big advantages of this method is that while it depends on domain knowledge, this does not require many various workshops with all relevant stakeholders but can be done in one quick session with people having general knowledge about process. Despite the complexity and variability of the procurement process, the method successfully identified critical activities and provided optimization recommendations that support improving procurement efficiency and standardization.
While leveraging Large Language Models (LLMs) for automated analysis and report generation offers innovative solutions, it is essential to have a Human in the Loop. Automated reports must be validated against the real experience of domain experts to ensure accuracy and relevance.

Future steps include creating a correlation matrix between different layers and their mapping to specific challenges, such as low sequential layer scores indicating potential issues in particular areas. Additionally, refining the architecture for automated report generation and adding features to support cooperation between analysts and LLMs will further enhance the effectiveness of the WISE methodology.

In summary, the WISE methodology is a valuable tool for organizations seeking to enhance their procurement processes, offering a structured approach to process improvement supported by advanced analytical capabilities.

\bibliographystyle{plain} 
\bibliography{refs}


\end{document}